\pdfoutput=1

\documentclass[11pt]{article}

\usepackage{ACL2023}

\usepackage{times}
\usepackage{latexsym}

\usepackage[T1]{fontenc}

\usepackage[utf8]{inputenc}

\usepackage{microtype}

\usepackage{inconsolata}

\usepackage{algorithm}
\usepackage{algpseudocode}
\usepackage{comment}
\usepackage{enumerate}
\usepackage{enumitem}
\usepackage{float}
\usepackage{pdfpages}

\usepackage{amsmath}
\usepackage{tabularx}
\usepackage{booktabs}
\usepackage{subcaption}
\usepackage{multirow}
\usepackage{amsfonts,amssymb,bbding,pifont,xcolor,makecell}
\newcommand{\cmark}{\ding{51}}%

\usepackage{newfloat}
\usepackage{listings}
\usepackage{graphicx}
\usepackage{xurl}
\usepackage{xparse}
\usepackage{array}
\usepackage{stfloats}
\usepackage{helvet}
\usepackage{courier}
\usepackage{colortbl}  
\usepackage[normalem]{ulem}

\newcommand{\blue}[1]{\textcolor{blue}{#1}}

\title{MIR-GAN: Refining Frame-Level Modality-Invariant Representations with Adversarial Network for Audio-Visual Speech Recognition}

\author{Yuchen Hu$^1$, Chen Chen$^1$, Ruizhe Li$^2$, Heqing Zou$^1$, Eng Siong Chng$^1$ \\
$^1$Nanyang Technological University, Singapore \quad$^2$University of Aberdeen, UK \\
\small{\texttt{\{yuchen005@e., chen1436@e., heqing001@e., aseschng@\}ntu.edu.sg,}} \\ \small{\texttt{ruizhe.li@abdn.ac.uk}}
}

\begin{document}
\maketitle
\begin{abstract}
Audio-visual speech recognition (AVSR) attracts a surge of research interest recently by leveraging multimodal signals to understand human speech.
Mainstream approaches addressing this task have developed sophisticated architectures and techniques for multi-modality fusion and representation learning.
However, the natural heterogeneity of different modalities causes distribution gap between their representations, making it challenging to fuse them.
In this paper, we aim to learn the shared representations across modalities to bridge their gap.
Different from existing similar methods on other multimodal tasks like sentiment analysis, we focus on the temporal contextual dependencies considering the sequence-to-sequence task setting of AVSR.
In particular, we propose an adversarial network to refine frame-level modality-invariant representations (MIR-GAN), which captures the commonality across modalities to ease the subsequent multimodal fusion process. 
Extensive experiments on public benchmarks LRS3 and LRS2 show that our approach outperforms the state-of-the-arts\footnote{Code is available at \url{https://github.com/YUCHEN005/MIR-GAN}.}.


\end{abstract}

\section{Introduction}
\label{sec:intro}
Human perception of the world intrinsically comprises multiple modalities, including vision, audio, text, etc.~\citep{mcgurk1976hearing,baltruvsaitis2018multimodal}.
Audio-visual speech recognition (AVSR) leverages both audio and visual modalities to understand human speech, improving the noise-robustness of audio-only speech recognition~\cite{chen2022noise,chen2022self,chen2023metric,chen2023unsupervised,hu2022interactive,hu2022dual,hu2023gradient,hu2023unifying,hu2023wav2code,zhu2023joint,zhu2023robust} with noise-invariant lip movement information~\citep{sumby1954visual}.
Thanks to recent advances of deep learning techniques, AVSR research has gained a remarkable progress~\citep{afouras2018deep,ma2021end,shi2022robust}.

Currently, the mainstream AVSR approaches are centered around developing sophisticated architectures and techniques for multi-modality fusion, including simple feature concatenation~\citep{makino2019recurrent,ma2021end,pan2022leveraging,chen2022leveraging,zhu2023vatlm}, recurrent neural network~\citep{petridis2018audio,xu2020discriminative} and cross-modal attention~\citep{afouras2018deep,lee2020audio,hu2023cross}.
Despite the advances, these approaches are often challenged by the representation gap persisting between naturally heterogeneous modalities~\citep{hazarika2020misa}.

\begin{figure}[t]
\centering
    \includegraphics[width=1.0\columnwidth]{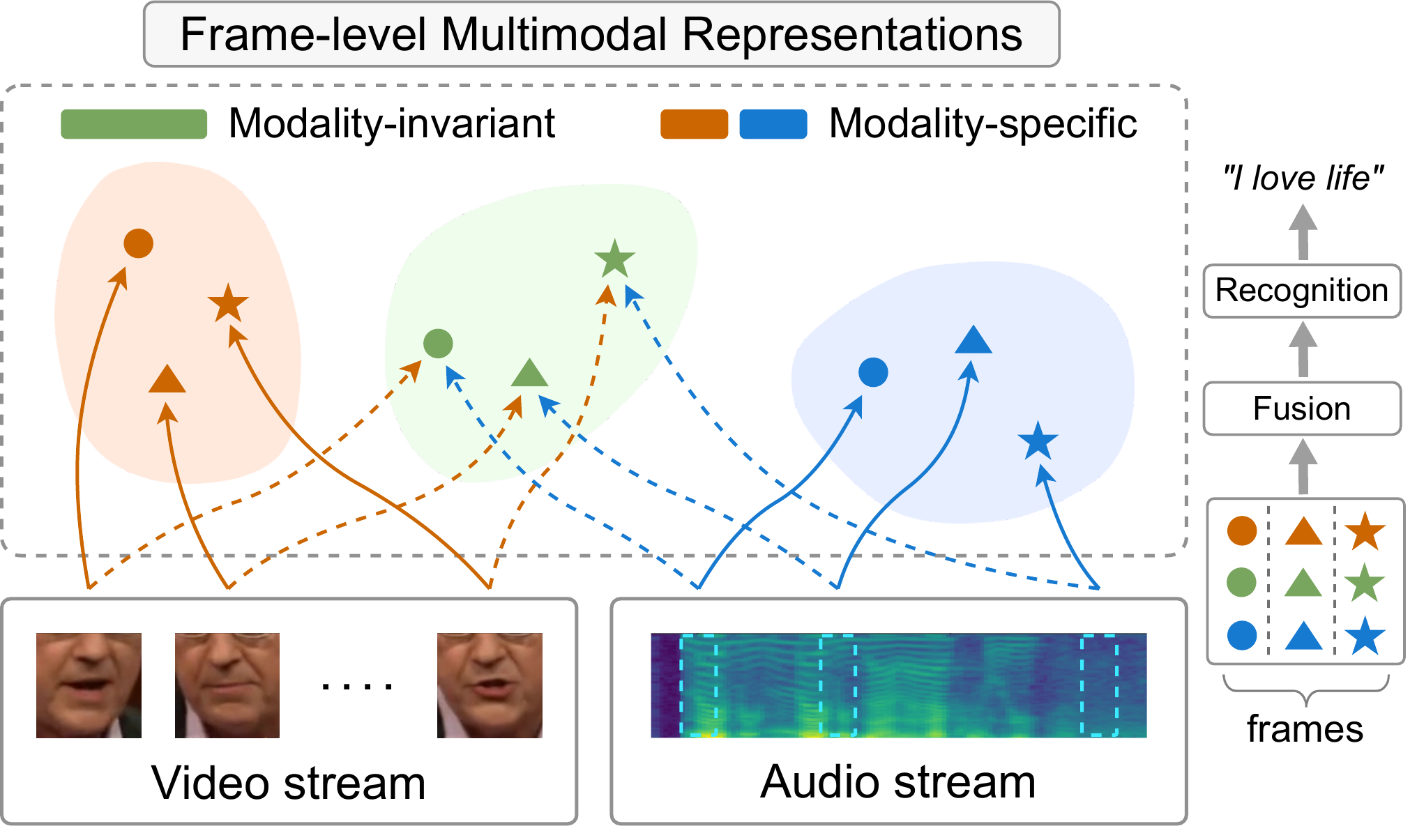}
    \vspace{-0.65cm}
    \caption{Multimodal learning of frame-level modality-invariant and -specific representations.}\label{fig1}
    \vspace{-0.45cm}
\end{figure}

Recently in some other multimodal tasks like sentiment analysis~\citep{hazarika2020misa,yu2021learning,yao2022modality} and cross-modal retrieval~\citep{xiong2020modality}, there have been research works proposing to learn two distinct representations to benefit multimodal learning.
The first representation is \emph{modality-invariant}, where multiple modalities of a same utterance are mapped to a shared space, indicating the homogeneous semantic meaning from the speaker.
In addition, they also learn \emph{modality-specific} representations that are private to each modality.
Given an utterance, each modality contains some unique features with respect to speaker-sensitive information~\citep{tsiros2013dimensions}.
Combing these two representations provides a holistic view of multimodal data for downstream tasks~\citep{yang2022learning}.
However, these methods focus on utterance-level representations that could be easily mapped to either shared or individual modality space using similarity cost functions, which does not apply to AVSR task that requires sequence-to-sequence mapping with temporal contextual dependencies~\citep{petridis2018audio}.

Motivated by above observations, we propose an adversarial network to refine frame-level modality-invariant representations (MIR-GAN) for capturing the commonality across modalities, which bridges their heterogeneous gap to ease the subsequent multimodal fusion.
In particular, we first design a MIR generator to learn modality-invariant representations over the shared audio-visual modality space.
Meanwhile, a modality discriminator is proposed to strengthen its modality agnosticism via adversarial learning.
Moreover, to further enrich its contextual semantic information, we propose a mutual information maximization strategy to align the refined representations to both audio and visual modality sequences.
Finally, both modality-invariant and -specific representations are fused for downstream speech recognition.
Empirical results demonstrate the effectiveness of our approach.
In summary, our main contributions are:
\begin{itemize}
\item We present MIR-GAN, an AVSR approach to refine frame-level modality-invariant representations, which captures the commonality across modalities and thus bridges their heterogeneous gap to ease multimodal fusion.

\item We first learn modality-invariant representations with a MIR generator, followed by another modality discriminator to strengthen its modality agnosticism via adversarial learning.
Furthermore, we propose a mutual information maximization strategy to enrich its contextual semantic information. 
Finally, both modality-invariant and -specific representations are fused for downstream recognition.

\item Our proposed MIR-GAN outperforms the state-of-the-arts on LRS3 and LRS2 benchmarks. Extensive experiments also show its superiority on ASR and VSR tasks.
\end{itemize}

\section{Related Work}
\label{sec:related_work}
\noindent\textbf{Audio-Visual Speech Recognition.}
Current mainstream AVSR methods focus on sophisticated architectures and techniques for audio-visual modality fusion.
Prior methods like RNN-T~\citep{makino2019recurrent}, Hyb-Conformer~\citep{ma2021end} and MoCo+wav2vec~\citep{pan2022leveraging} employ simple feature concatenation for multimodal fusion, other works including Hyb-RNN~\citep{petridis2018audio} and EG-seq2seq~\citep{xu2020discriminative} leverage recurrent neural network for audio-visual fusion.
In addition, cross-modal attention has also become popular recently for multimodal interaction and fusion in AVSR tasks, such as TM-seq2seq~\citep{afouras2018deep}, DCM~\citep{lee2020audio} and MMST~\citep{song2022multimodal}.
Despite the effectiveness, these fusion techniques are often challenged by the representation gap between naturally heterogeneous modalities.
Recently, multimodal self-supervised learning has been popular for capturing unified cross-modal representations, like AV-HuBERT~\citep{shi2022learning} and u-HuBERT~\citep{hsu2022u}, which achieve the state-of-the-art but require abundant unlabeled data and computing resources.
In this work, we propose a supervised learning scheme to efficiently refine modality-invariant representations for bridging the heterogeneous modality gap.

\noindent\textbf{Modality-Invariant and -Specific Representations.}
Recent studies in many multimodal tasks suggest that the model benefits from both shared and individual modality representations, including multimodal sentiment analysis~\citep{hazarika2020misa,yu2021learning,yang2022learning}, person re-identification~\citep{wei2021syncretic,huang2022exploring}, cross-modal retrival~\citep{zeng2022learning} and image-sentence matching~\citep{liu2019modality}, etc.
MISA~\citep{hazarika2020misa} maps the multimodal features into two spaces as modality-invariant and -specific representations, and then fuses them for downstream classification.
MCLNet~\citep{hao2021cross} learns modality-invariant representations by minimizing inter-modal discrepancy and maximizing cross-modal similarity.
VI-REID~\citep{feng2019learning} builds an individual network for each modality, with a shared identity loss to learn modality-invariant representations.
However, these methods map utterance-level representations to modality-invariant or -specific spaces via similarity cost functions, while AVSR is sequence-to-sequence task that requires contextual semantic information.
To this end, we propose an adversarial network with mutual information maximization to refine frame-level modality-invariant representations that subjects to temporal contextual dependencies.

\noindent\textbf{Adversarial Network.}
The concept of adversarial network starts from GAN~\citep{Goodfellow2014}, which has attracted a surge of research interests due to its strong ability of generating high-quality novel samples according to existing data.
The best-known applications include image-to-image translation~\citep{isola2017image} and image synthesis~\citep{denton2015deep,radford2015unsupervised}.
Recently, GAN is further applied to multimodal tasks such as text-to-image synthesis~\citep{reed2016generative,tan2020kt}, video captioning~\citep{yang2018video,bai2021discriminative} and cross-modal retrieval~\citep{qian2021dual}.
In this work, we leverage the strong distinguishing ability of adversarial network to strengthen the modality agnosticism of the learned modality-invariant representations.

\section{Methodology}
\label{sec:method}

\begin{figure*}[t]
\centering
    \includegraphics[width=0.96\textwidth]{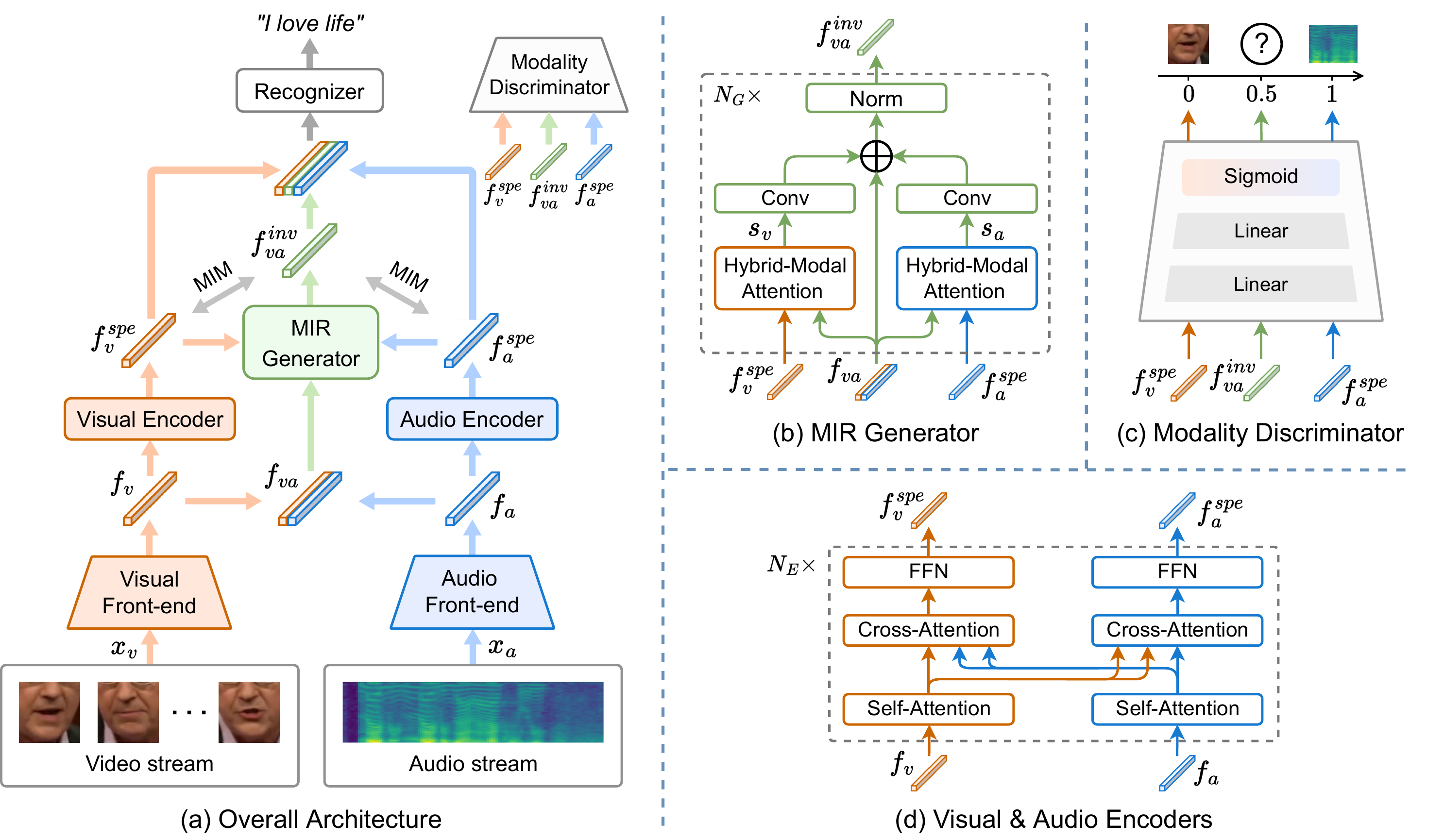}
    \vspace{-0.1cm}
    \caption{Illustration of our MIR-GAN. (a) Overall architecture. (b) MIR generator that learns modality-invariant representation $f^{inv}_{va}$. (c) Modality discriminator that strengthens the modality agnosticism of $f^{inv}_{va}$. (d) Visual and audio encoders that learn modality-specific representations $f^{spe}_{v}, f^{spe}_{a}$.
    ``MIM'' is mutual information maximization.}\label{fig2}
    \vspace{-0.3cm}
\end{figure*}

\subsection{Overview}
\label{ssec:overview}
The overall architecture of our proposed MIR-GAN is illustrated in Fig.~\ref{fig2}.
First, we have two front-end modules\footnote{Details are presented in Appendix~\ref{assec:model_setting}.} to process the input streams, which generate two modality sequences, \textit{i.e.}, $f_v, f_a \in \mathbb{R}^{T\times D}$, where $T$ is number of frames and $D$ is embedding size.
These two sequences are then fed by visual and audio encoders respectively to generate modality-specific representations, \textit{i.e.}, $f_v^{spe}, f_a^{spe} \in \mathbb{R}^{T\times D}$.
Based on that, we propose a MIR generator to learn modality-invariant representations by extracting the shared information of two modalities, \textit{i.e.}, $f_{va}^{inv} \in \mathbb{R}^{T\times D}$.
Meanwhile, we design a modality discriminator to strengthen its modality agnosticism via adversarial learning.
In addition, to further enrich its contextual semantic information, we propose a mutual information maximization (MIM) strategy to align the refined representations to both audio and visual modality sequences. 
Finally, both modality-invariant and -specific representations are fused for downstream speech recognition.

\subsection{Visual \& Audio Encoders}
\label{ssec:va_encoders}
As illustrated in Fig.~\ref{fig2} (d), we introduce a pair of visual and audio encoders to learn modality-specific representations.
Following Transformer~\citep{vaswani2017attention} architecture, they first employ self-attention modules to capture the contextual dependencies within each modality, followed by cross-attention modules for interaction between two modalities, which can initially narrow their gap to benefit the subsequent modality-invariant representation learning.
Finally, there are feed-forward networks to generate the modality-specific outputs.

\subsection{MIR-GAN}
\label{ssec:mir_gan}
With learned modality-specific representations, we propose MIR-GAN to refine frame-level modality-invariant representations.
First, we design a MIR generator to extract the shared information of two modalities, which generates a modality-invariant representation $f_{va}^{inv} \in \mathbb{R}^{T\times D}$.
Meanwhile, we design a modality discriminator to strengthen its modality agnosticism via adversarial learning.

\subsubsection{MIR Generator}
\label{sssec:mir_G}
Fig.~\ref{fig2} (b) details the architecture of proposed MIR generator $G$, where we design a hybrid-modal attention (HMA) module to extract out the part of information in each modality-specific representation that is related to both modalities:
\begin{equation}
\label{eq1}
\begin{split}
    s_m &= H\hspace{-0.04cm}M\hspace{-0.04cm}A(f_m^{spe}, f_{va}),\quad m \in \{v, a\},
\end{split}
\end{equation}
\noindent where the subscript $m$ denotes modality. The resulted features are then added to input sequence $f_{va}$ to form the final modality-invariant representation:
\begin{equation}
\label{eq2}
\begin{split}
    f_{va}^{inv} &= \text{Norm}(f_{va} + \sum_{m \in \{v, a\}}\text{Conv}(s_m)),
\end{split}
\end{equation}
\noindent where the ``Norm'' denotes layer normalization~\citep{ba2016layer}, ``Conv'' denotes 1$\times$1 convolution followed by PReLU activation~\citep{he2015delving}.

\begin{figure}[t]
\centering
    \includegraphics[width=0.62\columnwidth]{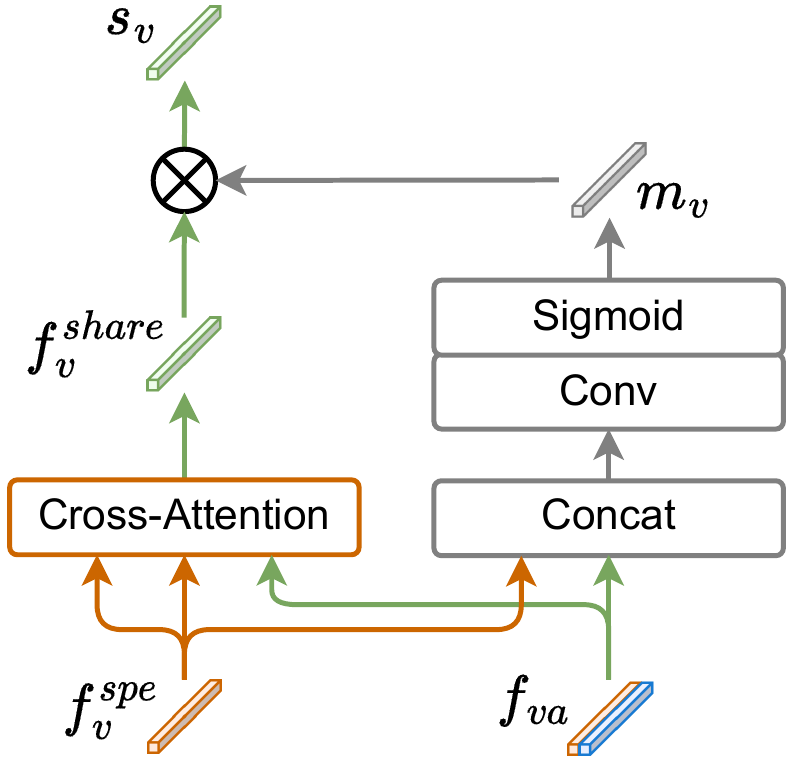}
    \vspace{-0.1cm}
    \caption{Illustration of the Hybrid-Modal Attention. Here we take the visual modality for example ($m = v$).}\label{fig3}
    \vspace{-0.2cm}
\end{figure}

\vspace{0.15cm}
\noindent\textbf{Hybrid-Modal Attention (HMA)} first involves a cross-attention sub-module to extract the information in each modality-specific representation that is related to both modalities, with the query input $f_{va}$ comprising both visual and audio sequence information, as shown in Fig.~\ref{fig3}:
\begin{equation}
\label{eq3}
\begin{split}
    f_{m}^{share} &= \text{Cross-Attention}(f_{va}, f_m^{spe}, f_m^{spe}),
\end{split}
\end{equation}

To further make the extracted feature invariant to modalities, we design a parallel convolutional network to learn a mask for filtering out the modality-specific information:
\begin{equation}
\label{eq4}
\begin{split}
    s_m &= f_{m}^{share} \otimes \sigma(\text{Conv}(f_m^{spe} \hspace{0.05cm}\Vert\hspace{0.05cm} f_{va})),
\end{split}
\end{equation}
\noindent where ``Conv'' denotes 1$\times$1 convolutional layer, $\Vert$ denotes feature concatenation, $\sigma$ denotes Sigmoid activation, $\otimes$ denotes element-wise multiplication.

As a result, the output representation $s_m$ from HMA involves information regarding both visual and audio modalities, making the final output $f_{va}^{inv}$ (in Eq.~\ref{eq2}) invariant to modalities.

\subsubsection{Modality Discriminator}
\label{sssec:modal_D}
With the generated modality-invariant representation, we further design a modality discriminator $D$ to strengthen its modality agnosticism via adversarial learning.
As shown in Fig.~\ref{fig2} (c), the discriminator consists of two linear layers followed by Sigmoid activation to predict a scalar between 0 and 1 for each frame, indicating which modality it belongs to (\textit{i.e.}, 0 for visual and 1 for audio):
\begin{equation}
\label{eq5}
\begin{split}
    D(f) \in \mathbb{R}^{T \times 1}, \hspace{0.2cm} f \in \{f_v^{spe}, f_a^{spe}, f_{va}^{inv}\},
\end{split}
\end{equation}

Therefore, for frames in modality-specific representations $f_v^{spe}$ and $f_a^{spe}$, we hope the discriminator can correctly classify the modality type, \textit{i.e.}, 0 or 1.
In contrast, in order to strengthen the modality agnosticism of refined representation $f_{va}^{inv}$, we hope it can confuse the discriminator with the output around 0.5, \textit{i.e.}, a medium between two modalities.

With above designs of generator and discriminator, the adversarial training objective of MIR-GAN can be mathematically formulated as:
\begin{equation}
\label{eq6}
\begin{split}
    \mathcal{L}_{G\hspace{-0.02cm}A\hspace{-0.02cm}N} &= \mathcal{L}_D + \mathcal{L}_{G} \\ 
    &= \mathbb{E}_f[\text{log}\hspace{0.02cm}D(f_a^{spe}) + \text{log}\hspace{0.02cm}(1 - D(f_v^{spe}))] \\
    &+ \mathbb{E}_f[-\text{log}\hspace{0.02cm}D(f_{va}^{inv}) - \text{log}\hspace{0.02cm}(1 - D(f_{va}^{inv}))],
\end{split}
\end{equation}
where $ f_{va}^{inv} = G(f_v^{spe}, f_a^{spe}, f_{va})$, $\mathbb{E}$ denotes the expectation over all the temporal frames in current data batch.
Details of the corresponding optimization strategy are illustrated in Alg.~\ref{alg1}.

\subsection{Mutual Information Maximization}
\label{ssec:mim}
The MIR-GAN successfully refines the modality-invariant representation by focusing on the modality commonality and agnosticism, while the original semantic information may not be preserved.
To this end, we further design a mutual information maximization (MIM) strategy via contrastive learning to enrich the contextual semantic information in refined modality-invariant representation.

In particular, we formulate a contrastive loss function to maximize the mutual information between modality-invariant representation $f_{va}^{inv}$ and the modality-specific representations $f_v^{spe}, f_a^{spe}$:

\begin{equation}
\label{eq7}
\begin{split}
    \mathcal{L}_{M\hspace{-0.03cm}I\hspace{-0.03cm}M} = &-\sum_{i=1}^{T}\text{log}\frac{\text{exp}(\langle f_{va\_i}^{inv}, f_{v\_i}^{spe} \rangle / \tau)}{\sum_{j=1}^{T} \text{exp}(\langle f_{va\_i}^{inv}, f_{v\_j}^{spe} \rangle / \tau)} \\
    &- \sum_{i=1}^{T}\text{log}\frac{\text{exp}(\langle f_{va\_i}^{inv}, f_{a\_i}^{spe} \rangle / \tau)}{\sum_{j=1}^{T} \text{exp}(\langle f_{va\_i}^{inv}, f_{a\_j}^{spe} \rangle / \tau)},
\end{split}
\end{equation}
where $\langle\hspace{0.04cm}\cdot, \cdot\hspace{0.04cm}\rangle$ denotes cosine similarity, $\tau$ is temperature parameter.
The subscripts $i$ and $j$ denote frame index, where $f_{va}^{inv}/f_v^{spe}/f_a^{spe} \in \mathbb{R}^{T\times D}$.

The constructed positive and negative samples are distinguished by frame index. 
As same frame of different representations express similar semantic meanings, we assign them as positive samples to strengthen consistency, while the mismatched frames are pulled apart from each other.
As a result, the MIM strategy can enrich the semantic information in final modality-invariant representation.

\subsection{Optimization}
\label{ssec:optim}

The optimization strategy of MIR-GAN is detailed in Alg.~\ref{alg1}.
After the forward-propagation process, we calculate $\mathcal{L}_{G\hspace{-0.02cm}A\hspace{-0.02cm}N}$ and $\mathcal{L}_{M\hspace{-0.03cm}I\hspace{-0.03cm}M}$ according to Eq.~\ref{eq6} and Eq.~\ref{eq7}.
Meanwhile, the downstream speech recognition loss $\mathcal{L}_{rec}$ is calculated as the cross-entropy between recognized text and the ground-truth transcription.
The final training objective of MIR-GAN can therefore be written as:
\begin{equation}
\label{eq8}
\begin{split}
    \mathcal{L} &= \mathcal{L}_{rec} + \lambda_{G\hspace{-0.02cm}A\hspace{-0.02cm}N} \cdot \mathcal{L}_{G\hspace{-0.02cm}A\hspace{-0.02cm}N} + \lambda_{M\hspace{-0.03cm}I\hspace{-0.03cm}M} \cdot \mathcal{L}_{M\hspace{-0.03cm}I\hspace{-0.03cm}M},
\end{split}
\end{equation}
where $\lambda_{G\hspace{-0.02cm}A\hspace{-0.02cm}N}, \lambda_{M\hspace{-0.03cm}I\hspace{-0.03cm}M}$ are weighting parameters to balance different training objectives.

Inspired by GAN training strategy~\citep{Goodfellow2014}, we split the back-propagation process into two steps.
First, we \emph{maximize} $\mathcal{L}_{G\hspace{-0.02cm}A\hspace{-0.02cm}N}$ to update the discriminator, where the generator is detached from optimization.
According to Eq.~\ref{eq6}, maximizing the first term of $\mathcal{L}_{G\hspace{-0.02cm}A\hspace{-0.02cm}N}$ (\textit{i.e.}, $\mathcal{L}_D$) trains the discriminator to correctly classify the two modalities, while increasing the second term amounts to informing discriminator that $f_{va}^{inv}$ is modality-specific and can either be visual or audio\footnote{Function $\text{log}(x) + \text{log}(1-x)$ reaches maximum at $x=0.5$, and the minimum is obtained around $x=0$ and $x=1$.\label{fn3}} (this is opposite to what we desire as modality-invariant).
Second, we freeze discriminator and update the rest network, where \emph{minimizing} $\mathcal{L}_G$ pushes the discrimination output of $f_{va}^{inv}$ to 0.5,\textsuperscript{\ref{fn3}} which is a medium between visual and audio modalities, \textit{i.e.}, modality-agnostic.
In addition, $\mathcal{L}_{rec}$ optimizes the downstream speech recognition model and $\mathcal{L}_{M\hspace{-0.03cm}I\hspace{-0.03cm}M}$ implements the MIM strategy.
The entire system is trained in an end-to-end manner with well-tuned weighting parameters.

\begin{algorithm}[t]
\caption{\small MIR-GAN Optimization.}
\label{alg1}
\small 
\begin{algorithmic}[1]
    \Require Training data $D$ that contains visual-audio pairs $(x_v, x_a)$ and the text transcription $y$. 
    The MIR-GAN network $\theta$ that consists of front-ends $\theta_{v\hspace{-0.02cm}f}$ and $\theta_{a\hspace{-0.02cm}f}$, encoders $\theta_{vae}$, MIR generator $\theta_G$, modality discriminator $\theta_D$ and downstream speech recognition model $\theta_{rec}$.
    Hyper-parameter weights $\lambda_{G\hspace{-0.02cm}A\hspace{-0.02cm}N}, \lambda_{M\hspace{-0.03cm}I\hspace{-0.03cm}M}$.
    \State Randomly initialize the entire system $\theta$.
    \If{select \emph{self-supervised setting}}
        \State Load the pre-trained AV-HuBERT for speech recognition model $\theta_{rec}$ and front-ends $\theta_{v\hspace{-0.02cm}f}, \theta_{a\hspace{-0.02cm}f}$
    \EndIf
    \While{not converged}
        \For{$(x_v, x_a) \in D$}
            \State \textsc{Forward-Propagation}:
            \State \hspace{0.3cm} $f_v = \theta_{v\hspace{-0.02cm}f}(x_v), f_a = \theta_{a\hspace{-0.02cm}f}(x_a)$ \Comment{\blue{front-ends}}
            \State \hspace{0.3cm} $f_v^{spe}, f_a^{spe} = \theta_{vae}(f_v, f_a)$ \Comment{\blue{encoders}}
            \State \hspace{0.3cm} $f_{va} = f_v\Vert f_a$
	        \State \hspace{0.3cm} $f_{va}^{inv} = \theta_G(f_v^{spe}, f_a^{spe}, f_{va})$ \Comment{\blue{Generator}}
	        \State \hspace{0.3cm} $\hat{y} = \theta_{rec}(f_v^{spe}\Vert f_a^{spe}\Vert f_{va}^{inv})$ \Comment{\blue{recognition}}
	        \State \textsc{Training objectives}:
	        \State \hspace{0.3cm} $\mathcal{L}_{G\hspace{-0.02cm}A\hspace{-0.02cm}N}$ ($\mathcal{L}_D$ and $\mathcal{L}_G$) in Eq.~\ref{eq6} \Comment{\blue{Discriminator}}
	        \State \hspace{0.3cm} $\mathcal{L}_{M\hspace{-0.03cm}I\hspace{-0.03cm}M}$ in Eq.~\ref{eq7} \Comment{\blue{MI maximization}}
	        \State \hspace{0.3cm} $\mathcal{L}_{rec} = \text{CrossEntropy}(\hat{y}, y)$
	        \State \textsc{Back-Propagation}: \Comment{\blue{adversarial training}}
	        \State \hspace{0.3cm} \textsc{Update discriminator}: \Comment{\blue{unfreeze $\theta_D$}}
	        \State \hspace{0.6cm} $\underset{\theta_D}{\arg\max}\hspace{0.1cm} \mathcal{L}_{G\hspace{-0.02cm}A\hspace{-0.02cm}N}$ 
	        \State \hspace{0.3cm} \textsc{Update the rest network}: \Comment{\blue{freeze $\theta_D$}}
	        \State \hspace{0.6cm} $\underset{\theta\backslash\theta_D}{\arg\min}\hspace{0.1cm} \mathcal{L}_{rec} + \lambda_{G\hspace{-0.02cm}A\hspace{-0.02cm}N} \cdot \mathcal{L}_G + \lambda_{M\hspace{-0.03cm}I\hspace{-0.03cm}M} \cdot \mathcal{L}_{M\hspace{-0.03cm}I\hspace{-0.03cm}M}$ 
        \EndFor
    \EndWhile
\end{algorithmic}
\normalsize
\end{algorithm}

\section{Experiments}
\label{sec:exp}

\subsection{Experimental Setup}
\label{ssec:setup}
\label{sssec:datasets}
\noindent\textbf{Datasets.} We conduct experiments on two large-scale public benchmarks, LRS3~\citep{afouras2018lrs3} and LRS2~\citep{chung2017lip}. 
LRS3 dataset collects 433 hours of transcribed English videos in TED and TEDx talks from over 5000 speakers, which is the largest publicly available labeled audio-visual speech recognition dataset. 
LRS2 dataset contains 224 hours of video speech, with a total of 144K clips from BBC programs.

\label{sssec:model_config}
\noindent\textbf{Model Configurations.} 
We first build a base model with only front-ends and downstream speech recognition module, which follows Transformer architecture with 24 encoder layers and 9 decoder layers.
Based on that, we build the MIR-GAN with $N_E=3$ visual \& audio encoder layers and $N_G=3$ MIR generator layers.
To maintain similar model size, we only use 12 encoder layers and 9 decoder layers in the recognition model.
The number of parameters in our base model and MIR-GAN are 476M and 469M respectively.
We also use Conformer~\citep{gulati2020conformer} as our backbone.
In addition, we implement a self-supervised setting by loading pre-trained AV-HuBERT\footnote{\url{https://github.com/facebookresearch/av_hubert}}.
Following prior work~\citep{shi2022robust}, we employ data augmentation and noisy test set based on MUSAN noise~\citep{snyder2015musan}.
More detailed settings are presented in Appendix~\ref{assec:model_setting} -~\ref{assec:train_detail}.

\label{sssec:baselines}
\noindent\textbf{Baselines.} 
To evaluate our proposed MIR-GAN, we select some popular AVSR methods for comparison, which can be roughly divided into two groups.
The first is supervised learning method, including TM-seq2seq/CTC~\citep{afouras2018deep}, RNN-T~\citep{makino2019recurrent}, EG-seq2seq~\citep{xu2020discriminative} and Hyb-Conformer~\citep{ma2021end}.
Another one is the recently popular self-supervised learning method such as MoCo+wav2vec~\citep{pan2022leveraging}, AV-HuBERT~\citep{shi2022robust} and u-HuBERT~\citep{hsu2022u}.

\begin{table*}[t]
\centering
\resizebox{0.95\textwidth}{!}{
\begin{tabular}{cccccccccc}
\toprule
\multicolumn{2}{c}{\multirow{2}{*}{Method}} & \multirow{2}{*}{Backbone} & \multirow{2}{*}{Criterion} & Unlabeled & Labeled & \multirow{2}{*}{DataAug} & \multirow{2}{*}{LM} & \multicolumn{2}{c}{WER(\%)} \\
\multicolumn{2}{c}{} & & & data (hrs) & data (hrs) & & & Clean & Noisy \\
\midrule
\multicolumn{10}{c}{\cellcolor[HTML]{EBEBEB} \emph{Supervised}}  \vspace{0.05cm}   \\
\multicolumn{2}{c}{TM-seq2seq~\citeyearpar{afouras2018deep}} & Transformer & S2S & - & 1,519 & \cmark & \cmark & 7.2 & - \\
\multicolumn{2}{c}{EG-seq2seq~\citeyearpar{xu2020discriminative}} & RNN & S2S & - & 590 & \cmark & - & 6.8 & - \\
\multicolumn{2}{c}{RNN-T~\citeyearpar{makino2019recurrent}} & RNN & RNN-T & - & 31,000 & - & - & 4.5 & - \\
\multicolumn{2}{c}{Hyb-Conformer~\citeyearpar{ma2021end}} & Conformer & S2S + CTC & - & 590 & \cmark & \cmark & \textbf{2.3} & - \\
\multicolumn{10}{c}{\cellcolor[HTML]{EBEBEB} \emph{Self-Supervised}}  \vspace{0.05cm}   \\
\multicolumn{2}{c}{AV-HuBERT~\citeyearpar{shi2022robust}} & Transformer & S2S & 1,759 & 433 & \cmark & - & 1.4 & \textbf{5.8}\\
\multicolumn{2}{c}{u-HuBERT~\citeyearpar{hsu2022u}} & Transformer & S2S & 2,211 & 433 & \cmark & - & \textbf{1.2} & - \\
\midrule
\multicolumn{10}{c}{\cellcolor[HTML]{EBEBEB} \emph{Proposed (Supervised)}}  \vspace{0.05cm}   \\
\multirow{4}{*}{Ours} & Base model & \multirow{2}{*}{Transformer} & \multirow{2}{*}{S2S} & \multirow{2}{*}{-} & \multirow{2}{*}{433} & \multirow{2}{*}{\cmark} & \multirow{2}{*}{-} & 3.5 & 14.8 \\
& MIR-GAN & & & & & & & 2.8 & 11.7\\
\cline{2-10}
& Base model & \multirow{2}{*}{Conformer} & \multirow{2}{*}{S2S} & \multirow{2}{*}{-} & \multirow{2}{*}{433} & \multirow{2}{*}{\cmark} & \multirow{2}{*}{-} & 2.5 & 10.9 \\
& MIR-GAN & & & & & & & \textbf{2.1} & \textbf{8.5} \\
\multicolumn{10}{c}{\cellcolor[HTML]{EBEBEB} \emph{Proposed (Self-Supervised)}}  \vspace{0.05cm}   \\
\multirow{2}{*}{Ours} & Base model & \multirow{2}{*}{Transformer} & \multirow{2}{*}{S2S} & \multirow{2}{*}{1,759} & \multirow{2}{*}{433} & \multirow{2}{*}{\cmark} & \multirow{2}{*}{-} & 1.4 & 5.8 \\
& MIR-GAN & & & & & & & \textbf{1.2} & \textbf{5.6} \\
\bottomrule
\end{tabular}
}
\caption{WER (\%) of our MIR-GAN and prior works on LRS3 benchmark.
``S2S'' denotes sequence-to-sequence loss~\citep{watanabe2017hybrid}, ``CTC'' denotes CTC loss~\citep{graves2006connectionist}, ``DataAug'' denotes noise augmentation, ``LM'' denotes language model rescoring. 
The noisy test set is synthesized using MUSAN noise~\citep{snyder2015musan}.}
\label{table1}
\vspace{-0.2cm}
\end{table*}

\begin{table}[t]
\centering
\vspace{0.05cm}
\resizebox{1.0\columnwidth}{!}{
\begin{tabular}{ccccc}
\toprule
\multicolumn{2}{c}{\multirow{2}{*}{Method}} & \multirow{2}{*}{Backbone} & \multicolumn{2}{c}{WER(\%)} \\
\multicolumn{2}{c}{} & & Clean & Noisy \\
\midrule
\multicolumn{5}{c}{\cellcolor[HTML]{EBEBEB} \emph{Supervised}}  \vspace{0.05cm}   \\
\multicolumn{2}{c}{TM-seq2seq~\citeyearpar{afouras2018deep}} & Transformer & 8.5 & - \\
\multicolumn{2}{c}{TM-CTC~\citeyearpar{afouras2018deep}} & Transformer & 8.2 & - \\
\multicolumn{2}{c}{Hyb-RNN~\citeyearpar{petridis2018audio}} & RNN & 7.0 & - \\
\multicolumn{2}{c}{LF-MMI TDNN~\citeyearpar{yu2020audio}} & TDNN & 5.9 & - \\
\multicolumn{2}{c}{Hyb-Conformer~\citeyearpar{ma2021end}} & Conformer & \textbf{3.7} & - \\
\multicolumn{5}{c}{\cellcolor[HTML]{EBEBEB} \emph{Self-Supervised}}  \vspace{0.05cm}   \\
\multicolumn{2}{c}{MoCo+wav2vec~\citeyearpar{pan2022leveraging}} & Transformer & \textbf{2.6} & - \\
\midrule
\multicolumn{5}{c}{\cellcolor[HTML]{EBEBEB} \emph{Proposed (Supervised)}}  \vspace{0.05cm}   \\
\multirow{4}{*}{Ours} & Base model & \multirow{2}{*}{Transformer} & 5.4 & 21.2 \\
& MIR-GAN & & 4.5 & 16.7 \\
\cline{2-5}
& Base model & \multirow{2}{*}{Conformer} & 3.9 & 15.8 \\
& MIR-GAN & & \textbf{3.2} & \textbf{11.9} \\
\multicolumn{5}{c}{\cellcolor[HTML]{EBEBEB} \emph{Proposed (Self-Supervised)}}  \vspace{0.05cm}   \\
\multirow{2}{*}{Ours} & Base model & \multirow{2}{*}{Transformer} & 2.3 & 7.3 \\
& MIR-GAN & & \textbf{2.2} & \textbf{7.0} \\
\bottomrule
\end{tabular}
}
\caption{WER (\%) of our MIR-GAN and prior works on the LRS2 benchmark.
Detailed configurations are further presented in Table~\ref{table6}.}
\label{table2}
\vspace{-0.55cm}
\end{table}

\subsection{Main Results}
\label{subsec:main_results}
We conduct experiments on two public datasets under \emph{supervised} and \emph{self-supervised} settings, depending on whether use the AV-HuBERT pre-trained model.
Results show that our proposed MIR-GAN achieves the state-of-the-art under both settings.

\noindent\textbf{LRS3 Benchmark.}
Table~\ref{table1} presents the AVSR performance of our proposed MIR-GAN and prior methods on LRS3 benchmark.
Under supervised setting, our MIR-GAN achieves significant improvement over the base model in both clean and noisy testing conditions, and the best performance achieves new state-of-the-art (2.1\% vs. 2.3\%) while without using the language model rescoring.
In addition, the Conformer backbone consistently outperforms Transformer (2.1\% vs. 2.8\%, 8.5\% vs. 11.7\%).
Under self-supervised setting, MIR-GAN also improves the performance of base model, which surpasses or matches previous state-of-the-art (1.2\% vs. 1.2\%, 5.6\% vs. 5.8\%) while using less unlabeled data for pre-training.

\noindent\textbf{LRS2 Benchmark.}
Table~\ref{table2} compares the AVSR results of MIR-GAN and baselines on LRS2 benchmark.
We can observe that the proposed MIR-GAN outperforms previous state-of-the-art by a large margin under both supervised and self-supervised settings (3.2\% vs. 3.7\%, 2.2\% vs. 2.6\%).
In addition, we also observe promising gains of performance in noisy testing conditions.

As a result, our proposed MIR-GAN achieves new state-of-the-art under both supervised and self-supervised settings on two public benchmarks, which demonstrates its superiority on AVSR task.

\begin{table*}[t!]
\centering
\resizebox{1.0\textwidth}{!}{
\begin{tabular}{c|cccccc|cccccc}
\toprule
\multirow{2}{*}{Model} & \multicolumn{2}{c}{TF-Sup-3} & \multicolumn{2}{c}{CF-Sup-3} & \multicolumn{2}{c|}{TF-SelfSup-3} & \multicolumn{2}{c}{TF-Sup-2} & \multicolumn{2}{c}{CF-Sup-2} & \multicolumn{2}{c}{TF-SelfSup-2} \\
 & Clean & Noisy & Clean & Noisy & Clean & Noisy & Clean & Noisy & Clean & Noisy & Clean & Noisy \\
\midrule
\textbf{MIR-GAN (Full)} & \textbf{2.8} & \textbf{11.7} & \textbf{2.1} & \textbf{8.5} & \textbf{1.2} & \textbf{5.6} & \textbf{4.5} & \textbf{16.7} & \textbf{3.2} & \textbf{11.9} & \textbf{2.2} & \textbf{7.0} \\
\midrule
\multicolumn{13}{c}{\cellcolor[HTML]{EBEBEB} \emph{Importance of Representations}}  \vspace{0.05cm}   \\
w/o Modality-Invariant & 3.3 & 13.7 & 2.4 & 10.1 & 1.3 & 5.8 & 5.3 & 19.9 & 3.7 & 14.9 & 2.3 & 7.2 \\
w/o Modality-Specific & 3.2 & 13.2 & 2.3 & 9.8 & 1.4 & 5.7 & 5.1 & 19.5 & 3.7 & 14.6 & 2.2 & 7.1 \\
\midrule
\multicolumn{13}{c}{\cellcolor[HTML]{EBEBEB} \emph{Importance of Modules}}  \vspace{0.05cm}   \\
w/o Visual \& Audio Encoders & 3.0 & 12.1 & 2.1 & 8.9 & 1.2 & 5.6 & 4.8 & 18.1 & 3.4 & 13.1 & 2.2 & 7.0 \\
w/o MIR Generator & 3.1 & 12.8 & 2.2 & 9.2 & 1.3 & 5.7 & 4.9 & 18.7 & 3.6 & 13.8 & 2.2 & 7.1 \\
w/o Modality Discriminator & 3.2 & 13.3 & 2.3 & 9.7 & 1.4 & 5.8 & 5.2 & 19.4 & 3.7 & 14.5 & 2.3 & 7.2 \\
\midrule
\multicolumn{13}{c}{\cellcolor[HTML]{EBEBEB} \emph{Importance of Strategies}}  \vspace{0.05cm}   \\
w/o Adversarial Training & 3.1 & 13.0 & 2.3 & 9.5 & 1.3 & 5.7 & 5.1 & 19.2 & 3.6 & 14.1 & 2.2 & 7.2 \\
w/o MIM Strategy & 2.9 & 12.0 & 2.1 & 9.0 & 1.2 & 5.6 & 4.7 & 17.8 & 3.5 & 12.6 & 2.2 & 7.1 \\
\bottomrule
\end{tabular}
}
\caption{Ablation study on LRS3 and LRS2 benchmarks.
Results are reported on six configurations in the format ``[Backbone]-[Setting]-[Test set]'', where ``TF''/``CF'' denote Transformer/Conformer backbone, ``Sup''/``SelfSup'' denote supervised/self-supervised setting, ``3''/2'' denote LRS3/LRS2 test set.}
\label{table3}
\vspace{-0.2cm}
\end{table*}

\subsection{Ablation Study}
\label{ssec:ablation}
Table~\ref{table3} presents the ablation study of each component in MIR-GAN.
There are three parts of ablation that are independent with each other, \textit{i.e.}, each study is conducted where other two components are kept same as the full MIR-GAN.

\vspace{0.05cm}
\noindent\textbf{Importance of Representations.}
We first investigate the importance of modality-invariant and -specific representations by discarding each of them.
When removing the refined modality-invariant representations from multi-modality fusion, the downstream speech recognition performance degrades a lot under all configurations, which verifies its significance of bridging the modality gap.
Similarly, we observe that the modality-specific representations also plays an important role in AVSR.

\vspace{0.05cm}
\noindent\textbf{Importance of Modules.}
In this part, we study the role of each module in the proposed MIR-GAN.
The visual and audio encoders are designed to extract deep modality-specific representations, which contributes to performance gains of MIR-GAN.
Then we replace the core module - MIR generator with simple feature concatenation in refining modality-invariant representations, which results in significant performance degradation.
Another key module - modality discriminator also contributes a lot in MIR-GAN by strengthening the modality agnosticism of refined representations from MIR generator.
In this sense, we conclude that all the modules in proposed MIR-GAN contribute positively to the multimodal representation learning.

\begin{figure}[t]
\centering
    \includegraphics[width=1.0\columnwidth]{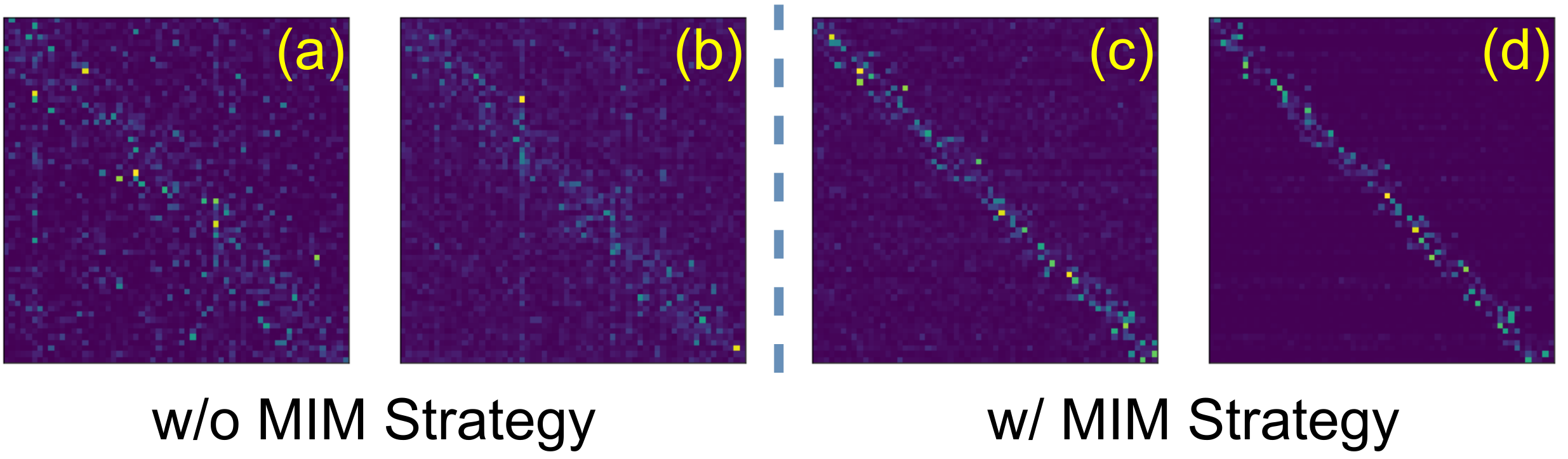}
    \vspace{-0.5cm}
    \caption{Alignment (attention map) between modality-invariant and -specific representations with and without MIM strategy: (a)(c) $f_{va}^{inv} \leftrightarrow f_v^{spe}$, (b)(d) $f_{va}^{inv} \leftrightarrow f_a^{spe}$.}\label{fig4}
    \vspace{-0.3cm}
\end{figure}

\begin{figure*}[t]
\centering
    \includegraphics[width=1.0\textwidth]{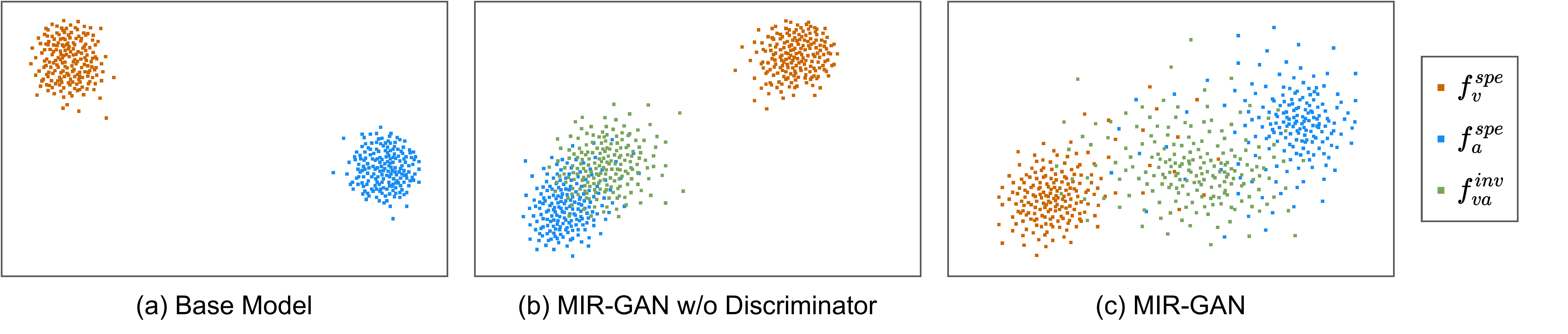}
    \vspace{-0.5cm}
    \caption{The t-SNE visualization of modality-invariant and -specific representations from (a) base model, (b) MIR-GAN without modality discriminator and (c) MIR-GAN. The orange and blue points denote visual and audio modality-specific representations respectively, and green points denote modality-invariant representations. This study is conducted on frame-level representations using a portion of LRS3 test set.}\label{fig5}
    \vspace{-0.2cm}
\end{figure*}

\vspace{0.05cm}
\noindent\textbf{Importance of Strategies.}
With the adversarial training strategy illustrated in Alg.~\ref{alg1}, the proposed modality discriminator effectively strengthens the modality agnosticism of the refined representations from generator.
To verify its effectiveness, we remove the adversarial training strategy from MIR-GAN, which results in similar performance degradation to the previous case without modality discriminator.
Therefore, it demonstrates the key role of this strategy in learning modality-invariant representations, where further visualization is shown in Fig.~\ref{fig5}.
Meanwhile, we design a MIM strategy to enrich the contextual semantic information in the refined modality-invariant representations, and similar performance drops can be observed in absence of such strategy.
Furthermore, we visualize the attention maps in Fig.~\ref{fig4} to show its effectiveness.
The clear diagonals in (c) and (d) indicate the strong ability of MIM strategy to align modality-invariant and -specific representations, which enriches the contextual semantic information in the former.

\vspace{0.05cm}
\noindent\textbf{Visualizations of Modality-Invariant and -Specific Representations.}
Fig.~\ref{fig5} presents the t-SNE visualization of modality-invariant and -specific representations to illustrate the principle of MIR-GAN.
First, we observe from (a) base model that the two modality-specific representations are distantly separated, indicating the heterogeneous gap between different modalities~\citep{hazarika2020misa}.
With the proposed MIR-GAN (no modality discriminator), the two modalities are pushed closer by the interaction between encoders, and the refined modality-invariant representations serve as a medium between them.
However, these refined representations are still entangled with audio modality-specific representations\footnote{Audio modality plays the dominant role in AVSR task.}, making them less modality-invariant.
Finally, the proposed discriminator effectively strengthens their modality agnosticism via adversarial learning, which are dispersed between two modalities to capture their commonality and thus bridge the heterogeneous modality gap.
As a result, the subsequent multi-modality fusion process would be eased and generate better features for downstream recognition.

\begin{table}[t]
\centering
\resizebox{1.0\columnwidth}{!}{
\begin{tabular}{p{16em}cc}
\toprule
\multirow{2}{*}{Method} & \multicolumn{2}{c}{WER(\%)} \\
& Clean & Noisy \\
\midrule
Base Model & 3.5 & 14.8 \\
\midrule
\quad + MCLNet~\citep{hao2021cross} & 3.4 & 14.5 \\
\quad + VI-REID~\citep{feng2019learning} & 3.3 & 14.0 \\
\quad + MISA~\citep{hazarika2020misa} & 3.3 & 13.7 \\
\midrule
MIR-GAN (ours) & \textbf{2.8} & \textbf{11.7} \\
\bottomrule
\end{tabular}
}
\vspace{-0.1cm}
\caption{Comparison between MIR-GAN and utterance-level multimodal approaches on LRS3 benchmark.}
\label{table4}
\vspace{-0.4cm}
\end{table}

\vspace{0.05cm}
\noindent\textbf{Comparison with Utterance-Level Approaches}
As illustrated in \S\ref{sec:related_work}, prior works have investigated utterance-level modality-invariant and -specific representations with similarity cost functions, including MISA~\citep{hazarika2020misa}, MCLNet~\citep{hao2021cross} and VI-REID~\citep{feng2019learning}.
We implement them in our framework as comparison to our proposed MIR-GAN, where we employ their designed similarity cost functions on frame-level representations.
As illustrated in Table~\ref{table4}, these utterance-level approaches can also improve AVSR results but still underperforms our proposed approach by a large margin.

\begin{table}[t]
\centering
\vspace{0.05cm}
\resizebox{1.0\columnwidth}{!}{
\begin{tabular}{cccccc}
\toprule
\multicolumn{2}{c}{\multirow{2}{*}{Method}} & \multirow{2}{*}{Backbone} & \multicolumn{3}{c}{WER(\%)} \\
\multicolumn{2}{c}{} & & AV & A & V \\
\midrule
\multicolumn{6}{c}{\cellcolor[HTML]{EBEBEB} \emph{Supervised}}  \vspace{0.05cm}   \\
\multicolumn{2}{c}{TM-seq2seq~\citeyearpar{afouras2018deep}} & Transformer & 7.2 & 8.3 & 58.9 \\
\multicolumn{2}{c}{EG-seq2seq~\citeyearpar{xu2020discriminative}} & RNN & 6.8 & 7.2 & 57.8 \\
\multicolumn{2}{c}{RNN-T~\citeyearpar{makino2019recurrent}} & RNN & 4.5 & 4.8 & \textbf{33.6} \\
\multicolumn{2}{c}{Hyb-Conformer~\citeyearpar{ma2021end}} & Conformer & \textbf{2.3} & \textbf{2.3} & 43.3 \\
\multicolumn{6}{c}{\cellcolor[HTML]{EBEBEB} \emph{Self-Supervised}}  \vspace{0.05cm}   \\
\multicolumn{2}{c}{Distill-Pretrain~\citeyearpar{ma2022visual}} & Conformer & - & - & 31.5 \\
\multicolumn{2}{c}{AV-HuBERT~\citeyearpar{shi2022robust}} & Transformer & 1.4 & 1.5 & \textbf{26.9} \\
\multicolumn{2}{c}{u-HuBERT~\citeyearpar{hsu2022u}} & Transformer & \textbf{1.2} & \textbf{1.4} & 27.2 \\
\midrule
\multicolumn{6}{c}{\cellcolor[HTML]{EBEBEB} \emph{Proposed (Supervised)}}  \vspace{0.05cm}   \\
\multirow{4}{*}{Ours} & Base model & \multirow{2}{*}{Transformer} & 3.5 & 4.7 & 63.5 \\
& MIR-GAN & & 2.8 & 3.5 & 48.6 \\
\cline{2-6}
& Base model & \multirow{2}{*}{Conformer} & 2.5 & 3.0 & 40.2 \\
& MIR-GAN & & \textbf{2.1} & \textbf{2.3} & \textbf{34.2} \\
\multicolumn{6}{c}{\cellcolor[HTML]{EBEBEB} \emph{Proposed (Self-Supervised)}}  \vspace{0.05cm}   \\
\multirow{2}{*}{Ours} & Base model & \multirow{2}{*}{Transformer} & 1.4 & 1.6 & 28.6 \\
& MIR-GAN & & \textbf{1.2} & \textbf{1.3} & \textbf{26.6} \\
\bottomrule
\end{tabular}
}
\vspace{-0.1cm}
\caption{Performance on single-modality inputs with LRS3 benchmark. 
``AV'', ``A'' and ``V'' indicate the input modality during both finetuning and inference stages.
The missing modality is replaced by zero embeddings.}
\label{table5}
\vspace{-0.4cm}
\end{table}

\vspace{0.05cm}
\noindent\textbf{Performance on Single-Modality Inputs.}
Furthermore, Table~\ref{table5} presents the performance of our MIR-GAN on single-modality inputs.
First, we observe that in all models using both modalities performs better than single modality, and the audio-only case achieves much better results than visual-only case, which shows the dominance of audio modality in AVSR task.
Under two single-modality cases, our proposed MIR-GAN both achieves significant improvement over the base model, and the best performance outperforms or matches previous state-of-the-arts in both supervised and self-supervised settings (2.3\% vs. 2.3\%, 34.2\% vs. 33.6\%; 1.3\% vs. 1.4\%, 26.6\% vs. 26.9\%).
Therefore, even with missing modality, our MIR-GAN can still refine effective modality-invariant representations to benefit the downstream speech recognition, which further verifies the generality of our approach.

\section{Conclusion}
\label{sec:conclusion}
In this paper, we propose MIR-GAN, an adversarial network to refine frame-level modality-invariant representations for AVSR, which captures the commonality across modalities to ease the multimodal fusion process.
MIR-GAN first learns modality-invariant representation with MIR generator, followed by a modality discriminator to strengthen its modality agnosticism via adversarial learning.
Furthermore, we propose a mutual information maximization strategy to enrich its contextual semantic information. 
Finally, both modality-invariant and -specific representations are fused to provide a holistic view of multimodal data for downstream task.
Experiments on public benchmarks show that our MIR-GAN achieves the state-of-the-art.

\section*{Limitations}
The main novelty of our proposed MIR-GAN is refining frame-level modality-invariant representations via adversarial learning.
It is promising to combine this approach with the popular self-supervised pre-training to learn unified multimodal representations.
In this work, we only load pre-trained AV-HuBERT for the front-ends and speech recognition model, while the proposed modules (\textit{i.e.}, encoders, generator, discriminator) are still trained from scratch.
In future, we may include the entire MIR-GAN into self-supervised learning scheme, together with the adversarial learning to refine better multimodal representations.

\section*{Ethics Statement}
All the data used in this paper are publicly available and are used under the following five licenses: the Creative Commons BY-NC-ND 4.0 License and Creative Commons Attribution 4.0 International License, the TED Terms of Use, the YouTube's Terms of Service, and the BBC’s Terms of Use.
The data is collected from TED and BBC and contain thousands of speakers from a wide range of races.
To protect the anonymity, only the mouth area of a speaker is visualized wherever used in the paper.

\section*{Acknowledgements}
The computational work for this article was partially performed on resources of the National Supercomputing Centre, Singapore (\url{https://www.nscc.sg}).


\bibliographystyle{acl_natbib}

\appendix

\section{Experimental Details}
\label{asec:exp_details}

\subsection{Datasets}
\label{subsec:datasets}
\noindent\textbf{LRS3}\footnote{\url{https://www.robots.ox.ac.uk/~vgg/data/lip_reading/lrs3.html}}~\citep{afouras2018lrs3} is currently the largest public sentence-level lip reading dataset, which contains over 400 hours of English video extracted from TED and TEDx talks on YouTube.
The training data is divided into two parts: pretrain (403 hours) and trainval (30 hours), and both of them are transcribed at sentence level.
The pretrain part differs from trainval in that the duration of its video clips are at a much wider range. 
Since there is no official development set provided, we randomly select 1,200 samples from trainval as validation set ($\sim$ 1 hour) for early stopping and hyper-parameter tuning.
In addition, it provides a standard test set (0.9 hours) for evaluation.

\vspace{0.05cm}
\noindent\textbf{LRS2}\footnote{\url{https://www.robots.ox.ac.uk/~vgg/data/lip_reading/lrs2.html}}~\citep{chung2017lip} is a large-scale publicly available labeled audio-visual (A-V) datasets, which consists of 224 hours of video clips from BBC programs.
The training data is divided into three parts: pretrain (195 hours), train (28 hours) and val (0.6 hours), which are all transcribed at sentence level.
An official test set (0.5 hours) is provided for evaluation use.

\begin{table*}[t!]
\centering
\resizebox{0.95\textwidth}{!}{
\begin{tabular}{cccccccccc}
\toprule
\multicolumn{2}{c}{\multirow{2}{*}{Method}} & \multirow{2}{*}{Backbone} & \multirow{2}{*}{Criterion} & Unlabeled & Labeled & \multirow{2}{*}{DataAug} & \multirow{2}{*}{LM} & \multicolumn{2}{c}{WER(\%)} \\
\multicolumn{2}{c}{} & & & data (hrs) & data (hrs) & & & Clean & Noisy \\
\midrule
\multicolumn{10}{c}{\cellcolor[HTML]{EBEBEB} \emph{Supervised}}  \vspace{0.05cm}   \\
\multicolumn{2}{c}{TM-seq2seq~\citeyearpar{afouras2018deep}} & Transformer & S2S & - & 1,519 & \cmark & \cmark & 8.5 & - \\
\multicolumn{2}{c}{TM-CTC~\citeyearpar{afouras2018deep}} & Transformer & CTC & - & 1,519 & \cmark & \cmark & 8.2 & - \\
\multicolumn{2}{c}{Hyb-RNN~\citeyearpar{petridis2018audio}} & RNN & S2S + CTC & - & 397 & \cmark & \cmark & 7.0 & - \\
\multicolumn{2}{c}{LF-MMI TDNN~\citeyearpar{yu2020audio}} & TDNN & LF-MMI & - & 224 & - & \cmark & 5.9 & - \\
\multicolumn{2}{c}{Hyb-Conformer~\citeyearpar{ma2021end}} & Conformer & S2S + CTC & - & 381 & \cmark & \cmark & \textbf{3.7} & - \\
\multicolumn{10}{c}{\cellcolor[HTML]{EBEBEB} \emph{Self-Supervised}}  \vspace{0.05cm}   \\
\multicolumn{2}{c}{MoCo+wav2vec~\citeyearpar{pan2022leveraging}} & Transformer & S2S + CTC & 60,000 & 381 & \cmark & - & \textbf{2.6} & - \\
\midrule
\multicolumn{10}{c}{\cellcolor[HTML]{EBEBEB} \emph{Proposed (Supervised)}}  \vspace{0.05cm}   \\
\multirow{4}{*}{Ours} & Base model & \multirow{2}{*}{Transformer} & \multirow{2}{*}{S2S} & \multirow{2}{*}{-} & \multirow{2}{*}{224} & \multirow{2}{*}{\cmark} & \multirow{2}{*}{-} & 5.4 & 21.2 \\
& MIR-GAN & & & & & & & 4.5 & 16.7 \\
\cline{2-10}
& Base model & \multirow{2}{*}{Conformer} & \multirow{2}{*}{S2S} & \multirow{2}{*}{-} & \multirow{2}{*}{224} & \multirow{2}{*}{\cmark} & \multirow{2}{*}{-} & 3.9 & 15.8 \\
& MIR-GAN & & & & & & & \textbf{3.2} & \textbf{11.9} \\
\multicolumn{10}{c}{\cellcolor[HTML]{EBEBEB} \emph{Proposed (Self-Supervised)}}  \vspace{0.05cm}   \\
\multirow{2}{*}{Ours} & Base model & \multirow{2}{*}{Transformer} & \multirow{2}{*}{S2S} & \multirow{2}{*}{1,759} & \multirow{2}{*}{224} & \multirow{2}{*}{\cmark} & \multirow{2}{*}{-} & 2.3 & 7.3 \\
& MIR-GAN & & & & & & & \textbf{2.2} & \textbf{7.0} \\
\bottomrule
\end{tabular}
}
\vspace{-0.1cm}
\caption{WER (\%) of our MIR-GAN and prior works on LRS2 benchmark.
``S2S'' denotes sequence-to-sequence loss~\citep{watanabe2017hybrid}, ``CTC'' denotes CTC loss~\citep{graves2006connectionist}, ``DataAug'' denotes noise augmentation, ``LM'' denotes language model rescoring. 
The noisy test set is synthesized using MUSAN noise~\citep{snyder2015musan}.}
\label{table6}
\vspace{-0.3cm}
\end{table*}

\subsection{Data Preprocessing}
\label{subsec:preprocessing}
The data preprocessing for above two datasets follows the LRS3 preprocessing steps in prior work~\citep{shi2022learning}.
For the audio stream, we extract the 26-dimensional log filter-bank feature at a stride of 10 ms from input raw waveform.
For the video clips, we detect the 68 facial keypoints using dlib toolkit~\citep{king2009dlib} and align the image frame to a reference face frame via affine transformation.
Then, we convert the image frame to gray-scale and crop a 96$\times$96 region-of-interest (ROI) centered on the detected mouth.
During training, we randomly crop a 88$\times$88 region from the whole ROI and flip it horizontally with a probability of 0.5. 
At inference time, the 88$\times$88 ROI is center cropped without horizontal flipping. 
To synchronize these two modalities, we stack each 4 neighboring acoustic frames to match the image frames that are sampled at 25Hz.

\subsection{Model Settings}
\label{assec:model_setting}
\noindent\textbf{Front-ends.} 
We introduce the modified ResNet-18 from prior work~\citep{shi2022learning} as visual front-end, where the first convolutional layer is replaced by a 3D convolutional layer with kernel size of 5$\times$7$\times$7. 
The visual feature is flattened into an 1D vector by spatial average pooling in the end.
For audio front-end, we use one linear projection layer followed by layer normalization~\citep{ba2016layer}.

\noindent\textbf{MIR-GAN.}
We build the MIR-GAN framework based on Transformer, where the embedding dimension/feed-forward dimension/attention heads in each Transformer layer are set to 1024/4096/16 respectively.
In addition, we also employ Conformer as backbone, where the depth-wise convolution kernel size is set to 31.
We use a dropout of $p = 0.1$ after the self-attention block within each Transformer layer, and each Transformer layer is dropped~\citep{fan2019reducing} at a rate of 0.1.

\subsection{Data Augmentation}
\label{assec:data_aug}
Following prior work~\citep{shi2022robust}, we use many noise categories for data augmentation.
We select the noise categories of ``\texttt{babble}'', ``\texttt{music}'' and ``\texttt{natural}'' from MUSAN noise dataset~\citep{snyder2015musan}, and extract some ``\texttt{speech}'' noise samples from LRS3 dataset.
All categories are divided into training, validation and test partitions.

During training process, we randomly select one noise category and sample a noise clip from its training partition. 
Then, we randomly mix the sampled noise with input clean audio, at signal-to-noise ratio (SNR) of 0dB with a probability of 0.25.

At inference time, we evaluate our model on clean and noisy test sets respectively.
Specifically, the system performance on each noise type is evaluated separately, where the testing noise clips are added at five different SNR levels: $\{-10, -5, 0, 5, 10\}dB$.
At last, the testing results on different noise types and SNR levels will be averaged to obtain the final noisy WER result.

\subsection{Training Details}
\label{assec:train_detail}
\noindent\textbf{Training.} 
We follow the sequence-to-sequence finetuning configurations of AV-HuBERT~\citep{shi2022robust} to train our systems.
We use Transformer decoder to decode the encoded features into unigram-based subword units~\citep{kudo2018subword}, where the vocabulary size is set to 1000.
The temperature $\tau$ in Eq.~\ref{eq7} is set to 0.1, and the weighting parameters $\lambda_{G\hspace{-0.02cm}A\hspace{-0.02cm}N}/\lambda_{M\hspace{-0.03cm}I\hspace{-0.03cm}M}$ in Eq.~\ref{eq8} are set to 0.01/0.005 respectively.
The entire system is trained for 60K steps using Adam optimizer~\citep{kingma2014adam}, where the learning rate is warmed up to a peak of 0.001 for the first 20K updates and then linearly decayed.
The finetuning process takes $\sim$ 1.4 days on 4 NVIDIA-V100-32GB GPUs.

\noindent\textbf{Inference.} No language model is used during inference.
We employ beam search for decoding, where the beam width and length penalty are set to 50 and 1 respectively.
All the hyper-parameters in our systems are tuned on validation set.
Since our experimental results are quite stable, a single run is performed for each reported result.

\subsection{Baselines}
\label{assec:baselines}
In this section, we describe the baselines for comparison.

\begin{itemize}
\item \textbf{TM-seq2seq}~\citep{afouras2018deep}: TM-seq2seq proposes a Transformer-based AVSR system to model the A-V features separately and then attentively fuse them for decoding, and uses sequence-to-sequence loss~\citep{watanabe2017hybrid} as training criterion.
\item \textbf{TM-CTC}~\citep{afouras2018deep}: TM-CTC shares the same architecture with TM-seq2seq, but uses CTC loss~\citep{graves2006connectionist} as training criterion.
\item \textbf{Hyb-RNN}~\citep{petridis2018audio}: Hyb-RNN proposes a RNN-based AVSR model with hybrid seq2seq/CTC loss~\citep{watanabe2017hybrid}, where the A-V features are encoded separately and then concatenated for decoding.
\item \textbf{RNN-T}~\citep{makino2019recurrent}: RNN-T adopts the popular recurrent neural network transducer~\citep{graves2012sequence} for AVSR task, where the audio and visual features are concatenated before fed into the encoder.
\item \textbf{EG-seq2seq}~\citep{xu2020discriminative}: EG-seq2seq builds a joint audio enhancement and multimodal speech recognition system based on the element-wise attention gated recurrent unit~\citep{zhang2019eleatt}, where the A-V features are concatenated before decoding.
\item \textbf{LF-MMI TDNN}~\citep{yu2020audio}: LF-MMI TDNN proposes a joint audio-visual speech separation and recognition system based on time-delay neural network (TDNN), where the A-V features are concatenated before fed into the recognition network.
\item \textbf{Hyb-Conformer}~\citep{ma2021end}: Hyb-Conformer proposes a Conformer-based~\citep{gulati2020conformer} AVSR system with hybrid seq2seq/CTC loss, where the A-V input streams are first encoded separately and then concatenated for decoding.
\item \textbf{MoCo+wav2vec}~\citep{pan2022leveraging}: MoCo+wav2vec employs self-supervised pre-trained audio and visual front-ends, \textit{i.e.}, wav2vec 2.0~\citep{baevski2020wav2vec} and MoCo v2~\citep{chen2020improved}, to generate better audio-visual features for fusion and decoding.
\item \textbf{AV-HuBERT}~\citep{shi2022learning, shi2022robust}: AV-HuBERT employs self-supervised learning to capture deep A-V contextual information, where the A-V features are masked and concatenated before fed into Transformer encoder to calculate masked-prediction loss for pre-training, and cross-entropy based sequence-to-sequence loss is used for finetuning.
\item \textbf{u-HuBERT}~\citep{hsu2022u}: u-HuBERT extends AV-HuBERT to a unified framework of audio-visual and audio-only pre-training.
\item \textbf{Distill-Pretrain}~\citep{ma2022visual}: Distill-Pretrain proposes a Conformer-based VSR framework with additional distillation from pre-trained ASR and VSR models.

\end{itemize}

\end{document}